# Patterns of population displacement during mega-fires in California detected using Facebook Disaster Maps

**Shenyue Jia[1], Seung Hee Kim[1], Son V. Nghiem[2], Paul Doherty[3], Menas Kafatos[1]**

[1] Center of Excellence in Earth Systems Modeling and Observations (CEESMO), Chapman University, Orange, California, USA
[2] NASA Jet Propulsion Laboratory, California Institute of Technology, Pasadena, California, USA
[3] National Alliance for Public Safety GIS (NAPSG) Foundation, Washington D.C., USA

E-mail: xxx@xxx.xx

## Abstract

Facebook Disaster Maps (FBDM) is the first platform providing analysis-ready population change products derived from crowdsourced data targeting disaster relief practices. We evaluate the representativeness of FBDM data using the Mann-Kendall test and emerging hot and cold spots in an anomaly analysis to reveal the trend, magnitude, and agglommeration of population displacement during the Mendocino Complex and Woolsey fires in California, USA. Our results show that the distribution of FBDM pre-crisis users fits well with the total population from different sources. Due to usage habits, the elder population is underrepresented in FBDM data. During the two mega-fires in California, FBDM data effectively captured the temporal change of population arising from the placing and lifting of evacuation orders. Coupled with monotonic trends, the fall and rise of cold and hot spots of population revealed the areas with the greatest population drop and potential places to house the displaced residents. A comparison between the Mendocino Complex and Woolsey fires indicates that a densely populated region can be evacuated faster than a scarcely populated one, possibly due to the better access to transportation. In sparsely populated fire-prone areas, resources should be prioritized to move people to shelters as the displaced residents do not have many alternative options, while their counterparts in densely populated areas can utilize their social connections to seek temporary stay at nearby locations during an evacuation. Integrated with an assessment on underrepresented communities, FBDM data and the derivatives can provide much needed information of near real-time population displacement for crisis response and disaster relief. As applications and data generation mature, FBDM will harness crowdsourced data and aid first responder decision-making.

Keywords: Facebook Disaster Maps, crowdsourced data, social media, Mann-Kendall trend, anomaly analysis, wildfires, California

## 1. Introduction

Abrupt population displacement is a common emergency response strategy applied before or during a natural disaster. Evaluating the effectiveness of this strategy and obtaining useful information for disaster relief during a large population displacement is of great interest to both policy-makers and researchers. The accessibility of cellular tower transmission data with geospatial signature has been greatly constrained amid the concern of data abuse and privacy preservation (1, 2). Therefore, many researchers and practitioners have turned to geolocation data that are crowdsourced through leading social media platforms to assimilate population movements in a timely manner and assist disaster relief. Crowdsourced geolocation data through location-based services are sensitive to the fluctuation of users' position change and respond promptly to the spatial movement and formation of clusters. For the most popular social media platforms like Twitter and Facebook, their crowdsourced geolocation data can serve as proxies of population distribution to answer the most basic yet important question for any crisis response task: "Where are the people?"

Studies and actual practices have been carried out using the geotagged Tweets and Flickr picture uploads to track and reveal the change and distribution of population during natural disasters and public safety crises (3-8). These successful applications also motivated the leading social media platforms to develop their official crisis response tools. Such efforts have greatly increased the popularity of crowdsourced data in time-sensitive situations for urgent emergency actions, with natural disasters being prime examples. While the representativeness of the social media population has been questioned (4, 9, 10), the steadily increasing pool of active users of these platforms (11) has helped capture a more accurate picture of human activities (12).

Current limitations to advance the use of crowdsourced data in emergency response can be framed into three major points. First, most studies using crowdsource data from social media, especially Twitter (5), were focused on algorithms and pipelines to effectively retrieve and clean the raw data from platforms but could not derive meaningful metrics to support decision-making (13-15). Second, pipelines of data processing applied by different teams sometimes varied, causing difficulties to compare the derived outcomes. Third, a knowledge detachment occurred between data analysts and first responders due to the different backgrounds and focus of interest (16).

These limitations can partly be alleviated by developing a universal and analysis-ready dataset of population movements for major disasters and emergencies across different parts of the world. Such a dataset can relieve analysts the complex pre-processing of crowdsourced data and support immediate analyses to decision-making as well as it can be compared across different situations. The launch of Disaster Maps by Facebook in 2017 is a major effort to address this need. As a part of Facebook Data for Good (dataforgood.fb.com) initiative, Facebook Disaster Maps (FBDM) tool provides population displacement information every 8 hours in a 1-km grid during any major natural disaster across the globe (17). Anonymized and aggregated based on the crowdsourced data from Facebook App and Facebook Safety Check App usage with location service enabled, FBDM not only provides population count changes over the period of natural disasters, but also generates metrics to capture population anomaly compared with a pre-crisis situation. In addition, FBDM offers network and power coverage maps to illustrate the Internet and power connectivity of users, based on the anonymized connection data through cellular network to Facebook server and the information of device charging from users (17). For the first time, an analysis-ready and near real-time population change product has become available to assist disaster relief works, including identifying the places of population aggregation during a disaster (18), retrieving the directions of population displacement (19), and evaluating the degree of resource needs during a disaster (20). Leveraging this dataset with precaution considerations of privacy preservation not only assists disaster relief works during particular disasters but also helps to identify disaster-prone communities and plans for effective mitigation strategies.

In this paper, we use the FBDM population change data to analyze the spatio-temporal patterns of population displacement during the Mendocino Complex fire and the Woolsey fire, two mega-fires which occurred in California, USA, in 2018. Our study first evaluates the representativeness of FBDM population, then calculates a series of derivatives from the basic FBDM population change data to reveal the spatio-temporal pattern of population change as well as hot spots. Furthermore, this study assesses the versatility of these derivatives by comparing the outcomes between mega-fires occurred in scarcely and densely populated regions.

## 2. Mendocino Complex Fire and Woolsey Fire

The Mendocino Complex (459,000 acres) and Woolsey fires (96,949 acres) were selected due to their exceptionally large perimeter and damage in California during the 2018 fire season, the first full-length fire season after the launch of FBDM. These two fires also represented the environmental and social settings of fire-prone areas in less populated, forest covered Northern California versus heavily populated, shrubland dominated Southern California.

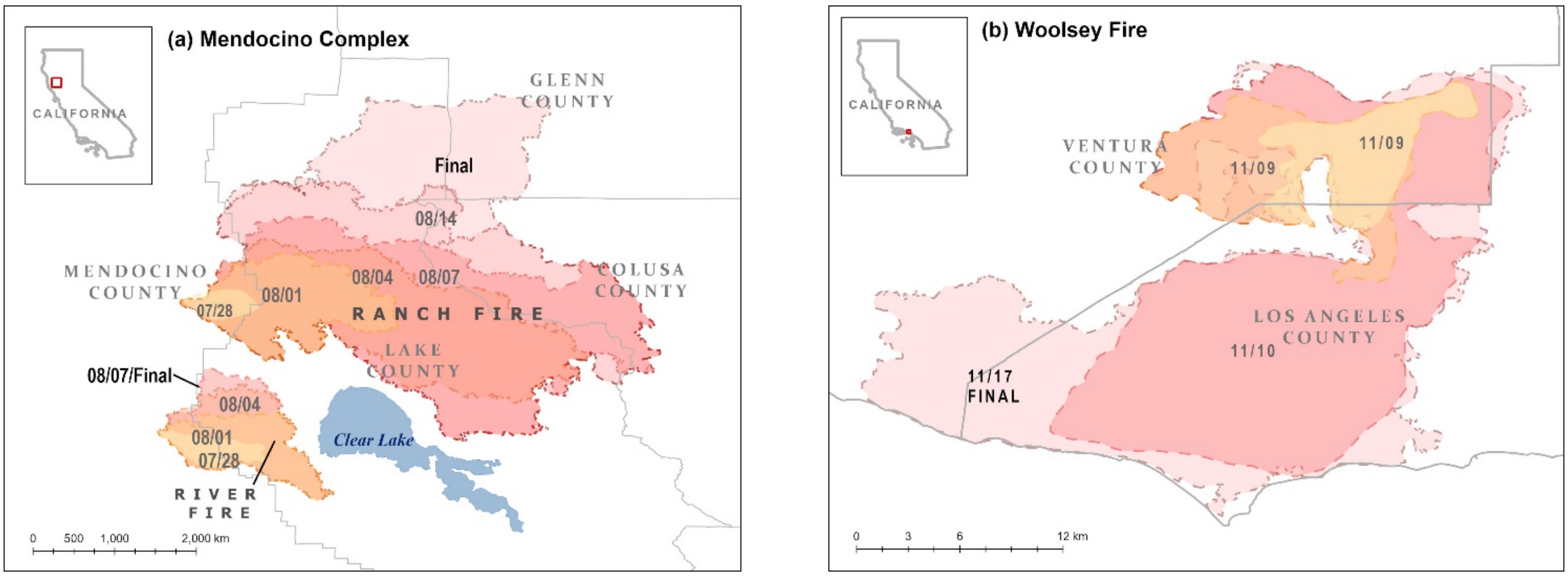

Figure 1. Progression and final perimeters of the Mendocino Complex fire (a) and the Woolsey fire (b). The change of fire perimeters as the burning continued is illustrated using boundaries with different colors and labels of dates. Note that maps are shown in different scales.

(a) Mendocino Complex fire

| | 7/27 | | | | 7/31 | 8/1 | | 8/3 | | | | 8/7 | 8/8 | 8/9 |
|---|---|---|---|---|---|---|---|---|---|---|---|---|---|---|
| River Fire | | | | | | | | | | | | | | |
| Ranch Fire Downwind East | | | | | | | | | | | | | | |
| Ranch Fire Downwind West | | | | | | | | | | | | | | |
| FBDM Data Availability | × | × | × | × | × | √ | √ | √ | √ | √ | √ | √ | √ | √ |

(b) Woolsey fire

| | 11/9 | 11/10 | | | 11/13 | | | | | | 11/19 |
|---|---|---|---|---|---|---|---|---|---|---|---|
| Near the origin of fire | | | | | | | | | | | |
| Outskirts of fire | | | | | | | | | | | |
| FBDM Data Availability | √ | √ | √ | √ | √ | √ | √ | √ | √ | √ | √ |

Evacuation Start — Evacuation Order Active — Evacuation Order lifted — √ FBDM Data Available — × FBDM Data Not Available

Figure 2. Timeline of fire progression and the placing/lifting of evacuation orders in and near the perimeters for Mendocino Complex fire (a) and Woolsey fire (b). Facebook Disaster Maps (FBDM) data availability is shown at the bottom of panel a and b.

Table 1. Basic information of Mendocino Complex fire and Woolsey fire (citation of data).

| **Fire name** | **Dates** | **Burned area** | **Land cover** | **Property destroyed** | **Fatalities** | **Population Density** |
|---|---|---|---|---|---|---|
| Mendocino Complex | Jul. 27-Nov. 7, 2018 | 1,858 $km^2$ | Temperate and boreal forest | 280 | 1 | 16 per $km^2$ |
| Woolsey | Nov. 8-Nov. 21, 2018 | 392 $km^2$ | Shrubland | 1643 | 3 | 335 per $km^2$ |

As the largest wildfire recorded in California's history, the Mendocino Complex fire started on July 27, 2018 and was not fully contained until November 7, 2018. This fire complex consists of two fires, the smaller, earlier contained River fire and the larger Ranch fire (Fig. 1a). Mandatory evacuations initiated and ended on different days near the perimeters of the River and Ranch fires (Fig. 2a). For the River fire, mandatory evacuation orders were placed shortly after its ignition on July 27 and lifted on July 31, which was before the situation escalated and the release of FBDM dataset. For the Ranch fire, mandatory orders were in place from August 3 to August 7 for the neighborhoods on the west side of the downwind direction. The burned area has a low population density at 76 people per $km^2$, thus causing less impacts than fires with similar size (Table 1).

In contrast, the Woolsey fire broke out in a heavily populated wildland urban interface (WUI) close to major metropolitan areas of Los Angeles and Ventura County in Southern California (Fig. 1b), and destroyed many more properties than the Mendocino Complex fire (Table 1). The fire quickly expanded southbound towards the Pacific coast by the strong Santa Ana winds over the shrubland. Mandatory evacuation orders were placed on November 8, 2018 for inland and coastal cities but were lifted at different times (Fig. 2b). The inland cities near the origin of fire lifted the mandatory evacuation orders on November 10, while cities located in the downward wind direction and the outskirts of fire perimeter kept the evacuation orders active until November 13 or later.

## 3. Data and Method

### *3.1 Data*

In this study, we focus on two key FBDM metrics, the Facebook baseline population (FBP) for pre-crisis situation and Facebook population z-score for anomalies during crisis. FBP is calculated based on the multi-day average login and usage of Facebook at a geographical location and a given time of the day over 5-13 weeks of the pre-crisis situation (17). Z-score is a standardized and unit-less metric derived to measure the departure of crisis population from the pre-crisis situation (18). Signs of z-score indicate the departure below (negative) or above (positive) the pre-crisis level. Z-score is comparable across the geographical location and time. A greater absolute value of z-score indicates a larger departure from the pre-crisis level. Both FBDM metrics are available every 8 hours during the major disasters and posted in 1-km grids to aggregate and anonymize the individual users (18). More information about FBDM data generation and accessibility is included in Supplemental Information Sec. 2.

In order to assess the representativeness of FBDM data, we employed two products of total population, the 2018 American Community Survey (ACS) population estimates at the zip code level from the U.S. Census as well as the Gridded Population of the World (GPW) version 4 (1-km grid) of the Columbia University (21).

We obtained the progression of fires and final fire perimeters from the USGS Geospatial Multi-Agency Coordination (GeoMAC) and the California Department of Forestry & Fire (CAL FIRE). Information of local government announcements on evacuation orders was from CAL FIRE and InciWeb (https://inciweb.nwcg.gov/). Geographical boundaries from U.S. Census TIGER were used to assist the analysis.

### *3.2 Calculating the penetration rate of Facebook users*

To assess the reliability of results drawn from FBDM data, we first evaluated the data representativeness using the penetration rate of Facebook service into the total population. Penetration rate was defined as below (Eq. 1).

$$Penetration\ Rate = \frac{Number\ of\ regular\ Facebook\ users}{Total\ population} \times 100\% \quad \text{(Eq. 1)}$$

In this study, we calculate two sets of penetration rates using different total population data, GPW in a 1-km grid and ACS population estimates at the zip-code level. For the penetration rate based on GPW, we resample the GPW data to the grid footprint of FBP to ensure the two datasets were co-registered. For the penetration rate based on ACS estimates, we identify FBDM grids inside each zip code and sum total the FBP values as the number of regular Facebook users. Penetration rates were calculated for all 36 time slices from the FBDM Woolsey fire dataset. We also calculate the zip code-level average for GPW-based penetration rate to better compare it with the ACS-based counterpart. We average penetration rate of all time slices and every time stamp to reveal the overall penetration level of Facebook usage and its diurnal cycle.

### *3.3 Deriving trends and emerging hotspots for Facebook population z-score*

We incorporated the space-time cube (22), a data structure designed to store spatial and temporal information to facilitate the analysis of population displacement pattern in space and time, including temporal trend, clustering of high and low values, and the evolution of the clusters. We construct the space-time cube by laying over the z-score map of every time stamp from the earliest to the latest.

Wildfire emergencies usually involve multiple critical events like the initiation and termination of mandatory evacuation orders. To reveal the change of population toward each of these events, we separate the z-score time series into sections based on the timing of these events and calculate the trend of z-score time series for each section using Mann-Kendall test, one of the widely used distribution-free tests of trend in time series analysis to reveal monotonic trend.(23-25). We also examine the temporal change of z-score over the entire space-time cube and detect the tipping points when the direction of trend changed. We then compare the detected tipping points with the official timing of the placing and lifting of evacuation orders. Such analysis is conducted for the z-score time series at each 1-km hexagon in the fire-affected area to reveal detailed spatial similarity or difference of population displacement, which can only be made possible with the crowdsourced data. If population change during the crisis presents as clusters in space, we calculate average z-score over each cluster to identify the intra-cluster similarity and inter-cluster distinction.

We conducted emerging hot spot analysis to detect trends in the clustering of z-score. The clustering is represented as hot and cold spots, or the aggregation of high and low population grids in space. During wildfires, such clustering in space might grow or shrink, indicating the dynamic of population displacement in space and time. To account for this process, we calculated the Getis-Ord Gi* statistics (26, 27) to detect hot and cold spots over the 1-km resolution z-score map at each time slice. Then we conduct the Mann-Kendall trend test for the Gi* statistics of all time slices at each 1-km grid. The Mann-Kendall trend of Gi* is calculated to categorize each 1-km grid as a new, unstable, or stable hot and cold spot, or no significant spatial aggregation. Such analysis replicates the Emerging Hot Spot Analysis from the Space Time Pattern Mining Toolbox of ArcGIS (28).

## 4. Results

### *4.1 Representativeness of Facebook Disaster Maps population data*

The penetration rate of Facebook usage varied by space. Regardless of the total population data source used in calculation, the penetration rate was higher in densely populated areas, especially along the major transportation corridors (Fig. 4). In less populated and remote areas near mountains and forests, the penetration rate of Facebook users was much lower. However, some spuriously high penetration grids also are present in remote areas (Fig. 4a) due to the noise in the original FBP data. Such an uncertainty needs to be considered while interpreting results derived for studies about wildfires, which often affect remote and less populated areas. In addition, areas with a higher proportion of younger population shows a higher penetration rate of Facebook users, indicating a sampling bias of the FBDM data (SI Fig. 1). The penetration rate of Facebook usage also indicates a diurnal cycle, which plunged after midnight and peaked in the mid-morning (SI Fig. 2).

FBP shows a high correlation with the total population. In the FBDM Woosley fire dataset, the $R^2$ between the average FBP and the total population from GPW and ACS 2018 estimates was 0.8322 and 0.8171, respectively (Fig. 3). The average penetration rate, shown as the slope of the linear fit equation in Fig. 3, indicated that FBP sampled 11.89% from the GPW and 7.91% from the ACS 2018 population estimates in the FBDM Woolsey fire dataset. GPW-based penetration rate was higher than the ACS-based rate, due to a systematical underestimation of population in GPW (29).

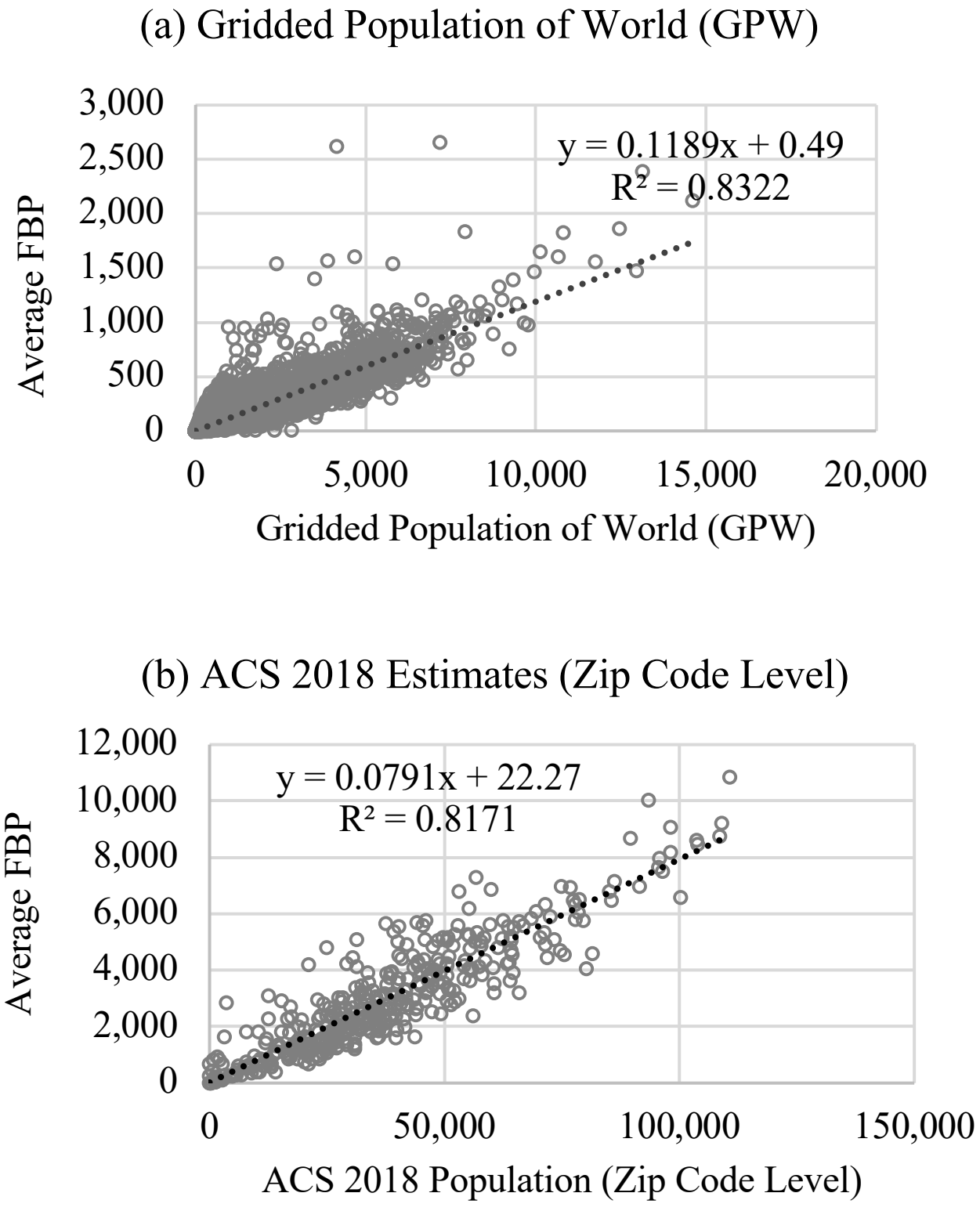

Figure 3. Average Facebook baseline population (FBP) from the FBDM Woolsey fire dataset and the total population. GPW 1 km population and ACS 2018 zip code level estimates were used in panel a and b as the total population

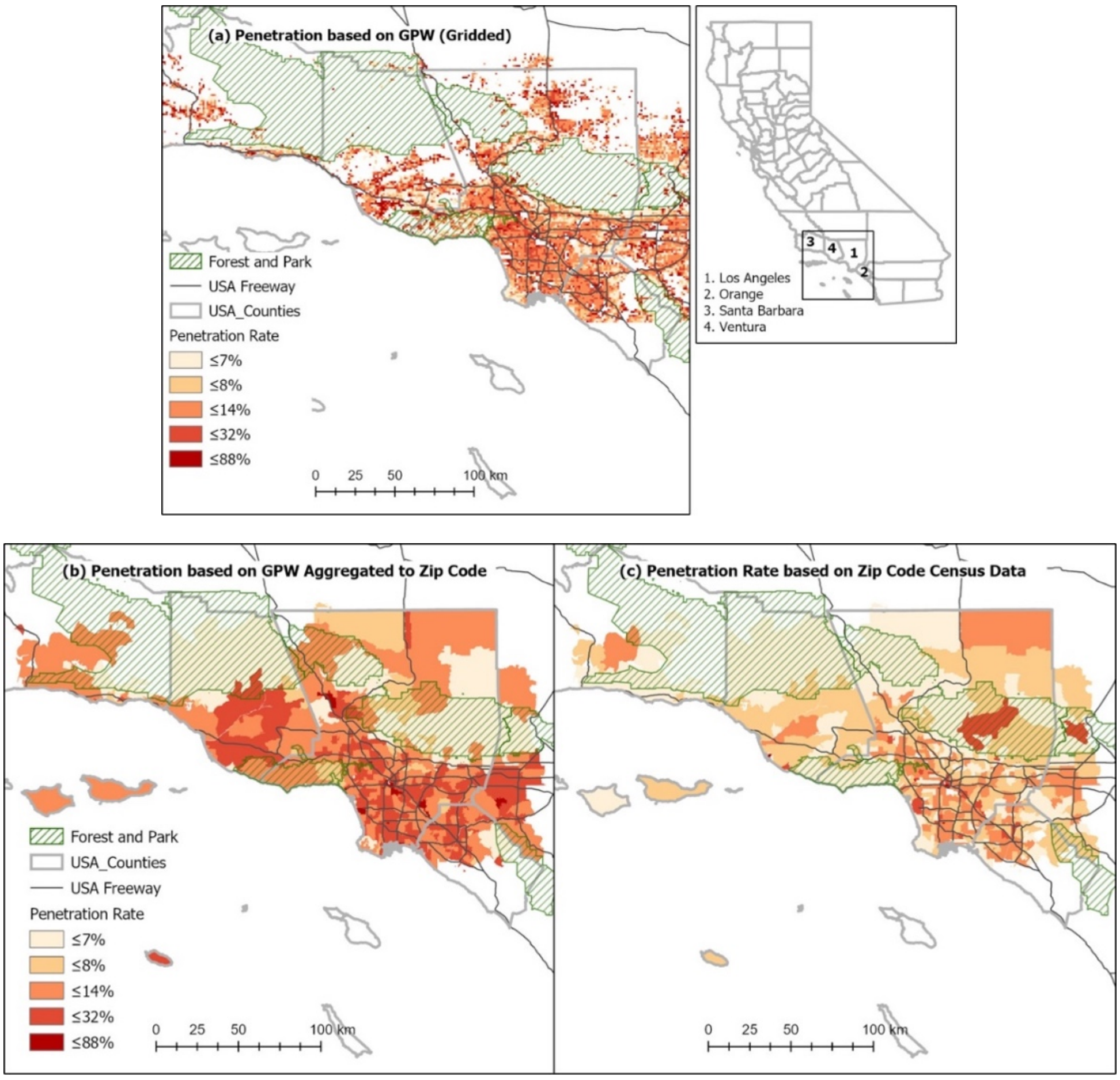

Figure 4. Spatial distribution of Facebook penetration rate based on the Facebook Disaster Maps (FBDM) Woolsey fire dataset. Panel a represents the penetration rate calculated based on GPW posted to 1 km grids. Panel b shows the zip code level average penetration rate based on GPW. Panel c shows the penetration rate calculated based on zip code level ACS 2018 population.

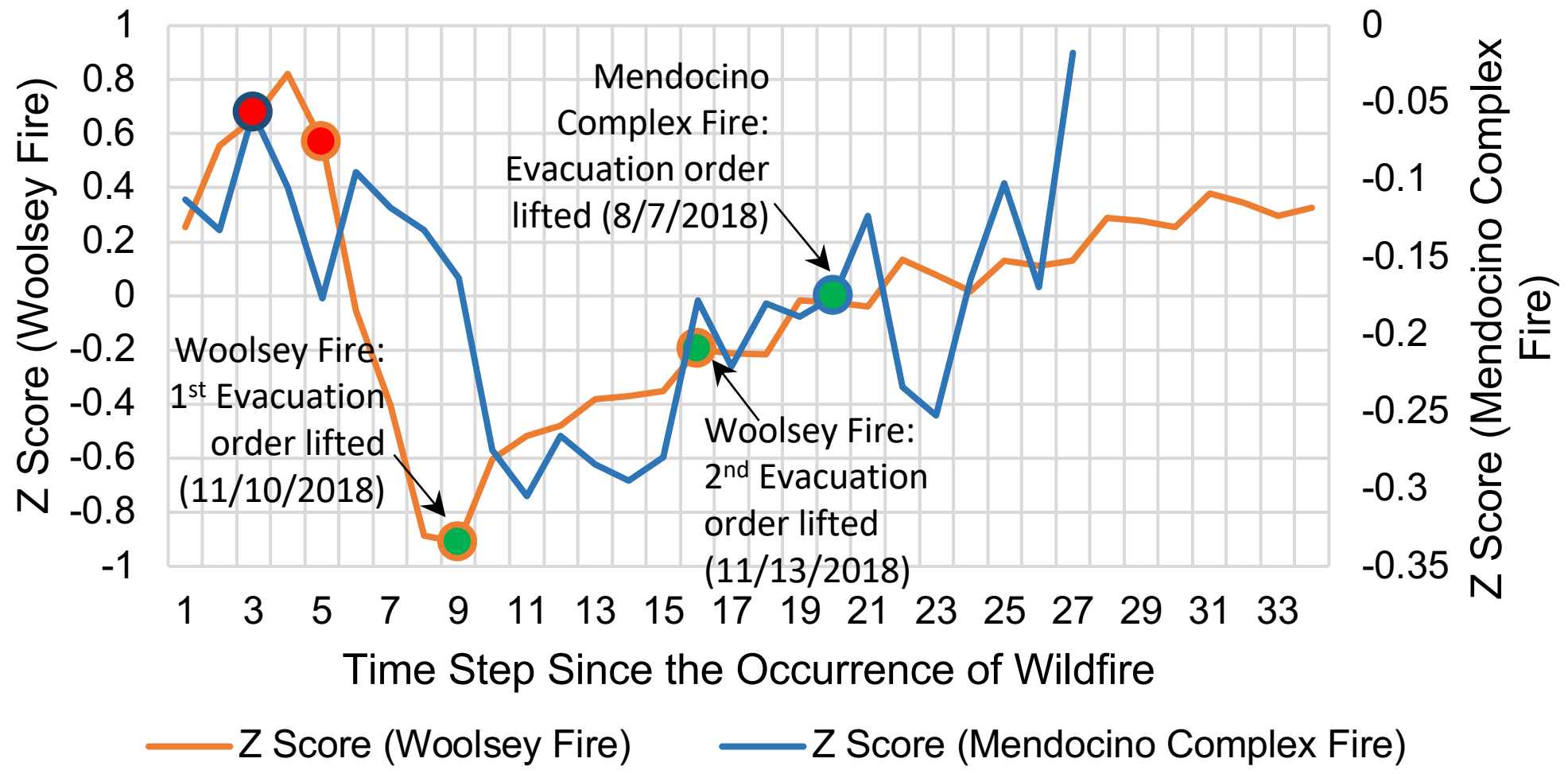

Figure 5. Facebook Disaster Maps (FBDM) z-score dynamics inside the entire region of interest affected by fire during the fire progression. Blue and orange lines represent the Mendocino Complex fire and the Woolsey fire, respectively. Red circles indicate the issue of an evacuation order. Green circles indicate evacuation order lifting. Z-scores are presented in an 8-hour time step. Positive z-score indicates that Facebook users are above the pre-crisis average, while negative z-score indicates that Facebook users are below the pre-crisis average.

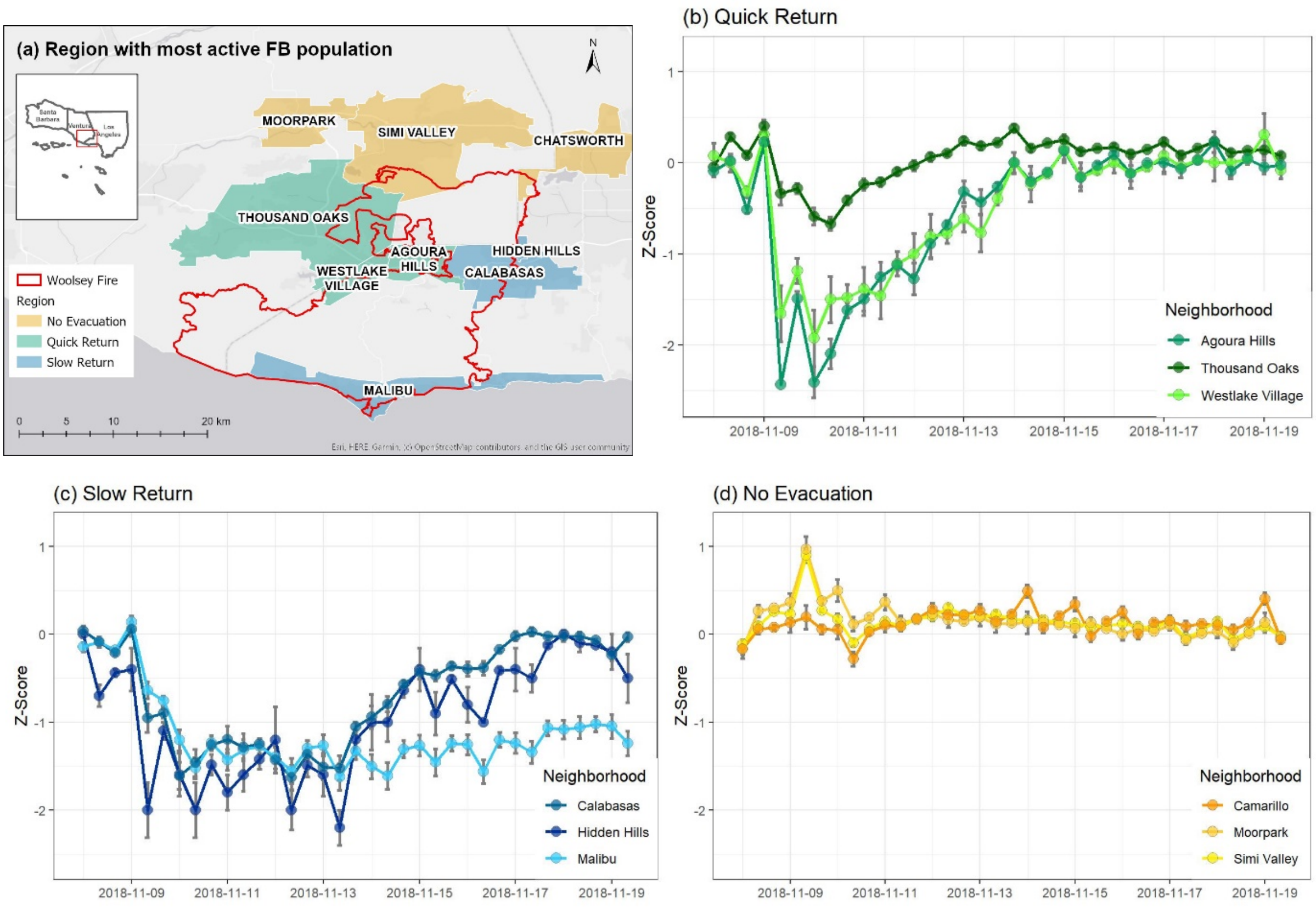

Figure 6. Facebook Disaster Maps (FBDM) z-score change of every 8 hours during the progression of Woolsey Fire for different locations due to the different fire progression extent. Green, blue, yellow indicated neighbourhoods with quick return, slow return, and no evacuation. Line charts from panel b to d represent the temporal change of z-score the three types of neighbourhoods mentioned above. Error bars are plotted using ± standard error. Positive z-score indicates that Facebook users are above pre-crisis level, while negative z-score indicates that Facebook users are below pre-crisis level.

### *4.2 Overall and regional population dynamics*

We found that the overall and regional average z-score captured the tipping points of population when communities responded to the placing and lifting of mandatory evacuation orders in both cases. The dynamics of z-score involved three stages: 1) drastic decrease after the evacuation order, with some significant increase near open shelters or nearby towns, 2) prominent increase after the lifting of evacuation order, and 3) slowing increase until z-score reached zero. The speed and magnitude of changes during each stage varied by location. During the Mendocino Complex fire, z-score took 9 time steps (i.e. 72 hours, time step 3 to 11) to decrease to its lowest since the placing of evacuation order (Fig. 5, blue line). For the Woolsey fire, such time was much shorter (32 hours, time step 5 to 9) (Fig. 5, orange line). The lowest z-score might stay for several time steps, depending on how soon the evacuation orders were lifted. Upon the lifting of evacuation orders, z-score climbed up in a slower pace than its decrease to reach zero (Fig. 5), representing overall situation of population distribution.

The average z-score also captured the spatial difference emerging from the differing time of evacuation orders placing and lifting by places. During the Woolsey fire, evacuation orders were lifted much sooner for neighborhoods near the fire origin than elsewhere. Such a difference resulted in three prominent types of population dynamic patterns in the fire-affected region, including area of quick return, slow return, and flash increase (Fig. 6). A subtle yet distinguishable difference within each type also indicates the linkage between the size of area affected and the magnitude of population movement. With a much smaller area and fraction of residents affected by the mandatory evacuation, Thousand Oaks showed a less drastic change of z-score than other neighborhoods in the quick return group, where the majority of the neighborhood was evacuated (Fig. 6b). A greater area exposed to wildfire progression and evacuation meant greater damage and longer road closure. Such inconveniences resulted in the low possibility and willingness to return, therefore led to a more gradual z-score recovery (Calabasas and Hidden Hills) or even no recovery at all (Malibu) (Fig. 6c). Aside from the decline of population, z-score also pinpointed possible destinations of the displaced people. Z-score increased slightly in November 9 outside the northern boundary of fire perimeters and quickly declined after November 10 (Fig. 6d), in parallel to the placing and lifting of the evacuation orders in Thousand Oaks and nearby neigbhorhoods (Fig. 6b). As the closest populated area not affected by the Woolsey fire in this region, these neighborhoods became potential destinations to accommodate the displaced people from Thousand Oaks and nearby places.

### *4.3 Grid-level trend in different stages of wildfire emergencies*

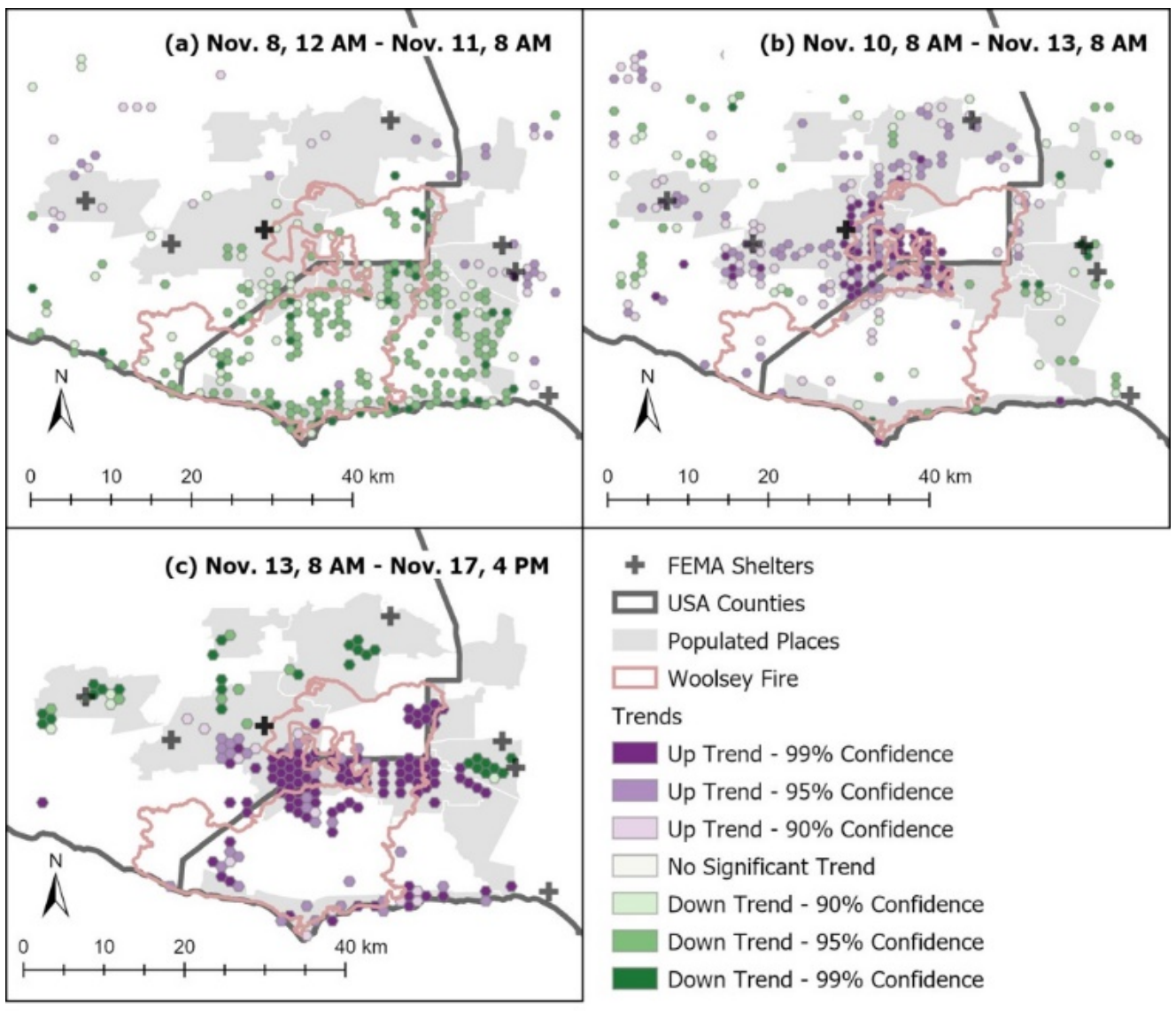

Figure 7. Significant trend of z-score in and adjacent to the Woolsey fire perimeter when communities responded to evacuation placing and lifting. Panels illustrate three stages during the fire emergency: (a) during the significant population drop after the placing of evacuation order, (b) while the evacuation orders were active or partly lifted, and (c) after evacuation orders were completely lifted (c).

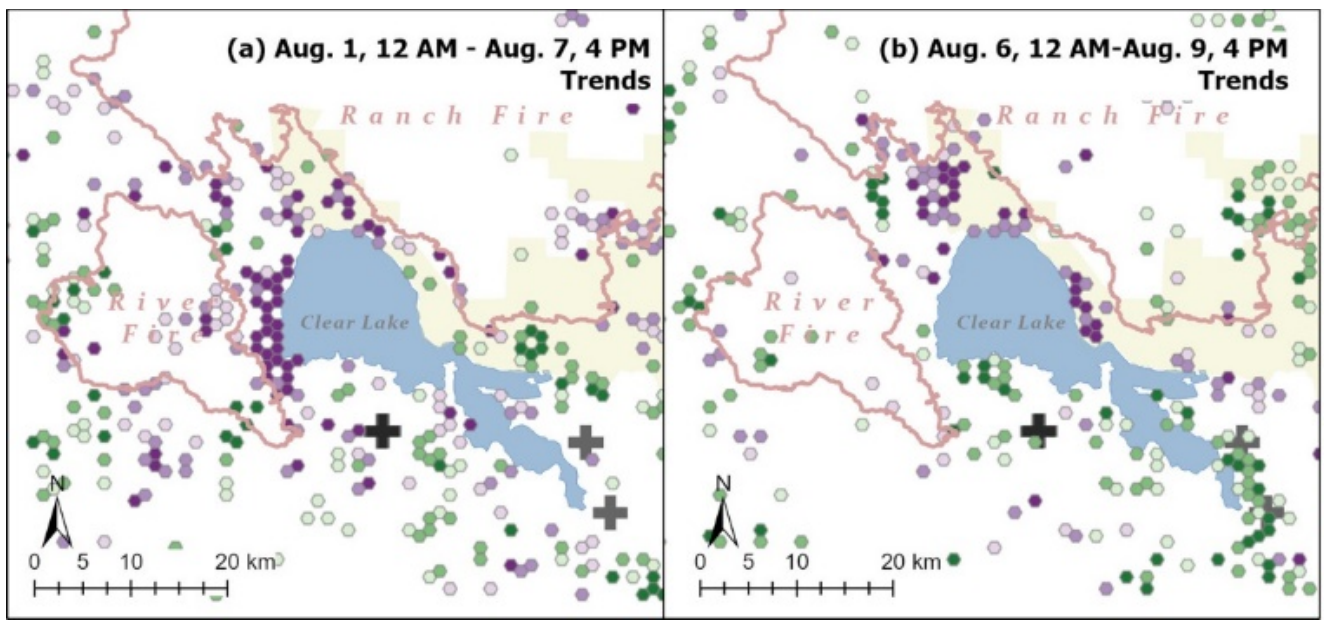

Figure 8. Significant trend of z-score in and adjacent to the Mendocino Complex fire perimeter when communities responded to evacuation placing and lifting. Panels illustrate two stages during the fire emergency: (a) while the evacuation orders were active or partly lifted and (b) after evacuation orders were completely lifted. Legend is the same as Fig. 7.

Trends of z-score at each grid showed a consistent pattern with the overall and regional average values. During the Woolsey fire, trends showed a more homogenouos pattern in space as the evacuation initiated concurrently across the region and terminated simoutanously within nearby neighborhoods. Most areas affected by the Woolsey fire showed a significant decrease of z-score while people were fleeing the evacuated area, with scattered increasing grids near some open shelters (Fig. 7a). When the lifting of evacuation order started, the spatial aggregation of increased z-score was prominent (Fig. 7b). Such increase expanded as the lifting continued (Fig. 7c). In contrast, open shelters showed a significant decrease as the lifting of evacuation continued (Fig. 7b-c). For the Mendocino Complex fire, the trend of z-score also captured the population changes in areas undergoing different types of population displacement. As evacuation orders to the west of the River fire perimeter were lifted before the FBDM data became available, this area showed significant increase of z-score while areas to the south of the Ranch fire was undergoing a population decrease after the newly placed evacuation order (Fig. 8a). When the evacuation orders near Ranch fire were lifted, positive trend occurred in parallel to the z-score decrease near the two OPEN shelters in the southeast corner of Clear Lake (Fig. 8b). Overall, the trend straightforwardly showed the most active areas of population decrease and increase during wildfire emergencies, providing insights on population displacement.

### *4.4 Changing hot and cold spots of population displacement*

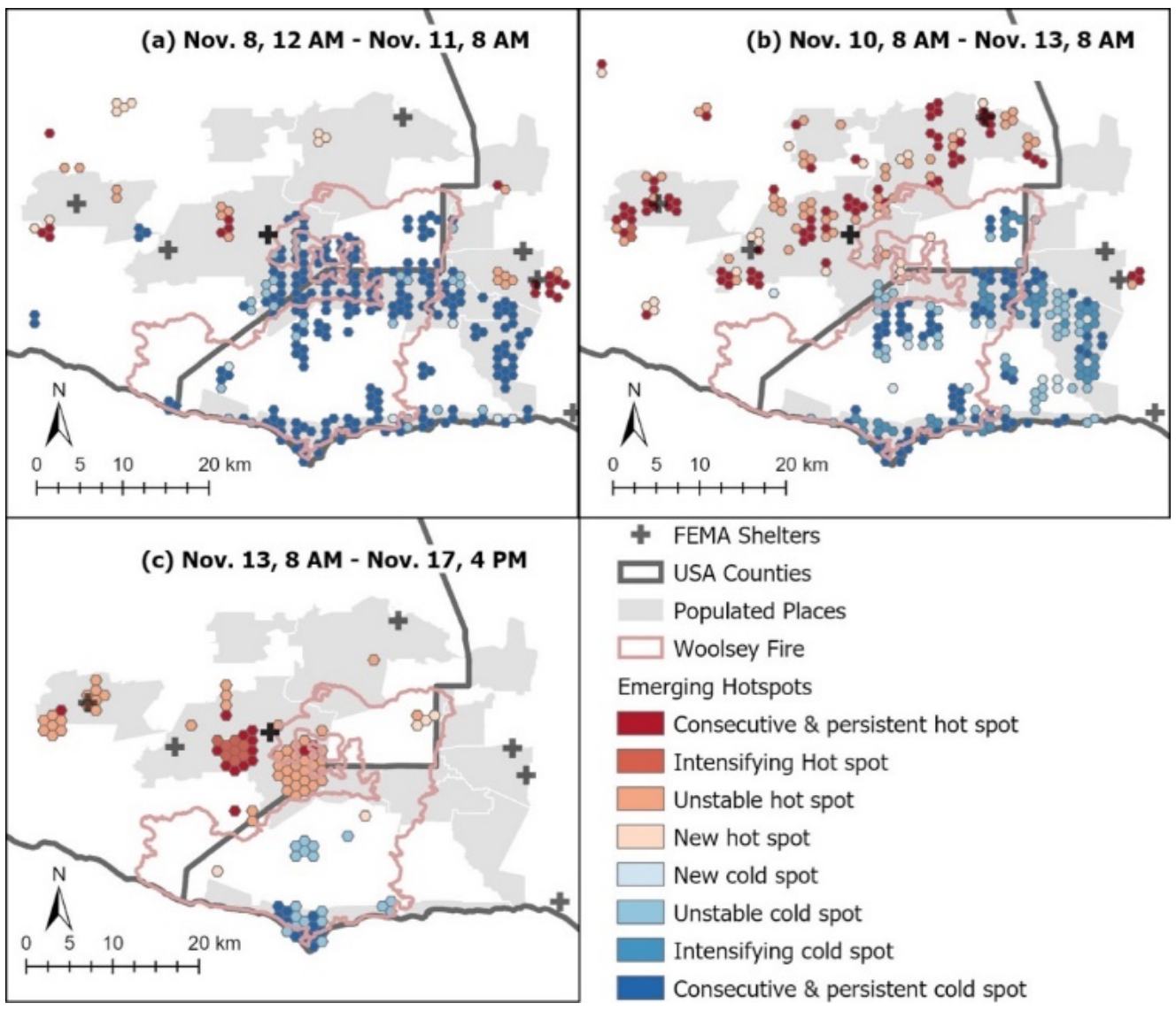

Figure 9. Emerging hot and cold spots of the Woolsey fire when communities responded to evacuation placing and lifting. Panels illustrate three stages during the fire emergency: (a) during the significant population drop after the announcement of evacuation order, (b) while the evacuation

orders were active or partly lifted, and (c) after evacuation orders were completely lifted.

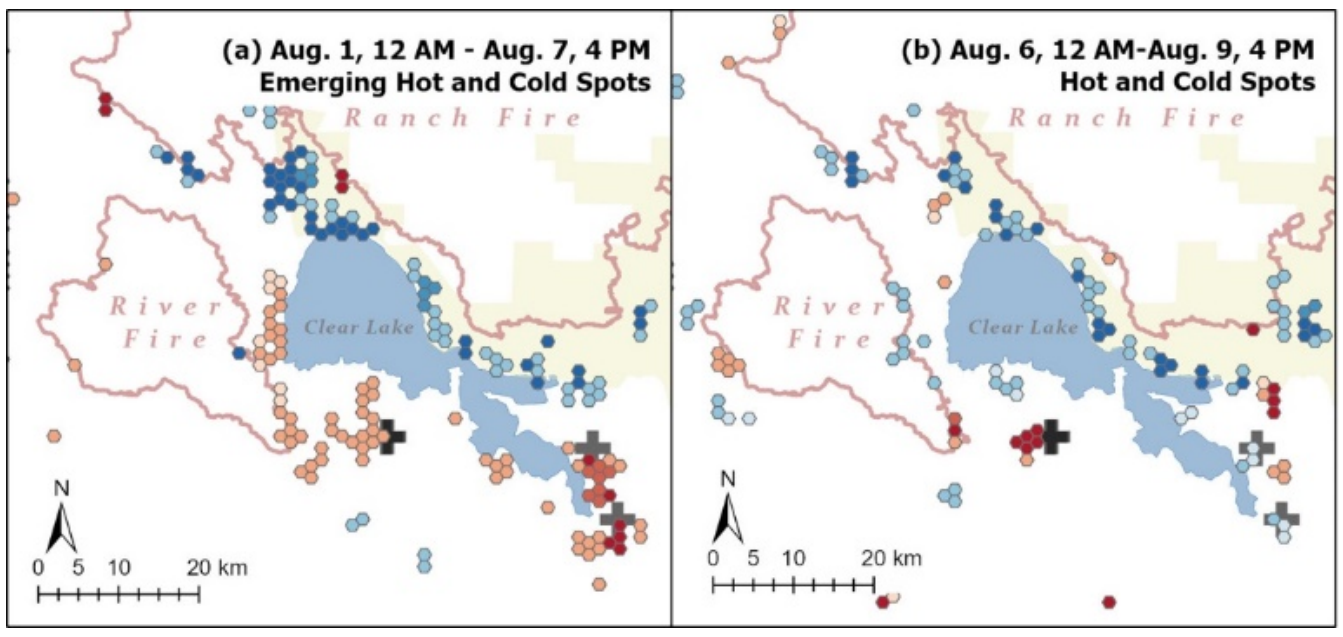

Figure 10. Emerging hot and cold spots of the Mendocino Complex fire when communities responded to evacuation placing and lifting. Panels illustrate two stages during the fire emergency: (a) while the evacuation orders were active or partly lifted and (b) after evacuation orders were completely lifted. Legend is the same as Fig. 9.

As spatial clustering of population anomaly, hot and cold spots of z-score might expand, persist, shrink, or disappear as a wildfire emergency unfolds. Recording the rise and fall of them not only documents the degree of population displacement but also identifies the temporal places of the population. During an active evacuation period, cold spots were consecutive and persistent across the evacuated area, with new cold spots popping up near its outskirt (Fig. 9a, 10a). When the evacuation orders were lifted, cold spots either became less stable or disappeared (Fig. 9b, 10b), which was consistent with the increasing trend of z-score (Fig. 7b, 8b). When most residents returned home, a large number of cold or hot spots became unstable or even disappeared (Fig. 7c, 8b), indicating the end of the anomalous population change arising from the displacement. Compared to cold spots, hot spots were less common while the population displacement was active. Yet, they represented possible places where displaced people were housed, providing crucial information for disaster relief works. A series of stable hot spots arose at or near the open shelters and towns not affected by the fire while the evacuation orders were in place (Fig. 9a-b and 10a). New and unstable hot spots appeared in areas experiencing a strong wave of population return after evecuation ended (Fig. 9c, Fig. 10a near the River fire). Meanwhile, some consecutive and persistent hot spots near open shelters disappeared as people housed in these facilities and neighborhoods left for home (Fig. 10b). The rise and fall of hot spots near open shelters was more prominent during the Mendocino Complex fire, partly due to the fewer options people have in this less populated, relatively isolated region in Northern California.

## 5 Discussion and Conclusions

FBDM data could reveal the trend, magnitude, and spatial clustering of population displacement in a timely manner with adequate representativeness out of the total population in area with intensive usage such as California. Though biased in age and other demographic factors, FBDM promptly captured the abrupt population displacement when the placing and lifting of mandatory evacuation orders occured. The dataset was sensitive not only to the prominent decrease of population in the evacuated area but also the significant and flashy increase of population resulted from the displacement. The timely response from the platform and the provision of comparable and analysis-ready products provides a foundation for trend analysis and spatial pattern detection to support decision making. State and local agencies design information products that could be used for monitoring local events like wildfires. We believe FBDM can improve usability of the platform by increasing the frequency at which data are released and by partnering with organizations who can produce actionable information products.

Distilling map series of population into meaningful secondary derivatives was necessary to harness the power of FBDM data. In this study, we calculated the Mann-Kendall trend and derived emerging hot and cold spots of Facebook population z-score to reveal the spatio-temporal pattern of population displacement. The rise of persistent and intensifying hot spots near open shelters during the evacuation was crucial information for disaster relief practice and post-disaster strategy. The analytical pipeline is highly transferrable to other disasters or emergencies with a better predictability, such as tornadoes, hurricanes and planned power outage amid extreme wildfire risk.

Comparability, a crucial feature of FBDM data, is critical for post-disaster reflection and future planning improvements. The same protocol of data generation enables the intercomparison between fires occurring in regions with varying social and environmental settings. In this study, we found that the evacuation in densely populated fire-prone area was faster due to the size of geographical area involved and the connectivity to transportation network. Population density affects the needs of resources and a less populated region might need resources prioritized to open shelters. In contrast, open shelters were not identified as hot spots in densely populated regions, as the displaced residents may have alternative options for temporary stay by exploiting their social connections (30) in nearby communities.

FBDM under-samples the elder population due to the relatively low popularity of Facebook in these age groups. Although such bias does not significantly affect the detection of the overall displacement trend and pattern, it could overlook the communities demanding more assistance yet failing to voice themselves out in the FBDM data. An analysis of social vulnerability based on the latest demographic data would be necessary (31) to identify the underrepresented but highly vulnerable communities. Such a discrepancies should be taken into consideration while using FBDM data to evaluate whether a neighborhood has been sufficiently evacuated and advising first responders to target the most vulnerable neighborhoods. In addition, the Internet connectivity can be disrupted during emergency situations, leading to an underestimation of human activities (32). The connectivity of mobile network map from FBDM (17) should be employed to account for such uncertainty and identify places with a higher possibility for a false signal of population patterns.

Challenging the so-called "The Tyranny of the Tweet" (33) in a crisis response sector, FBDM is the first platform to provide analysis-ready products from crowdsourced data for disaster relief in a timely manner. We have found that FBDM population z-score is a useful metric to reveal the temporal trend and spatial clustering during the massive population displacement arising from major disasters such as mega-fires. Derivatives from z-score, such as the Mann-Kendall trend, hot/cold spots, and the change of hot/cold spots can provide insightful information to evaluate the progress of population evacuation and return, identify agglomeration of displaced population, and assess the effectiveness of disaster relief strategy. With an uncertainty evaluation on the underrepresented groups, an improved workflow, and a coupled analysis with auxiliary products (movement vectors, mobile connectivity), FBDM population can help first responders deliver assistance to much needed communities and assist decision-makers to mitigate the disorder and chaos during crisis response. As illustrated in this paper, the FBDM method is versatile and thus adaptable to extreme wildfire situations such as the recent bushfires in Australia that have caused record-breaking population displacement (34). In short, future research should focus on providing FBDM in "the right format, to the right people, at the right time" in order to allow decision-makers and first responders to enhance outocmes for survivors.

Regarding future research, together with FBDM or similar crowdsourced data and independent remote sensing observations, the method presented in this paper can be adapted for applications to disasters in assessing the efficacy of emergency measures, such as the unprecedented massive change of population movement during hurricanes or quarantine or lockdown orders amid the pandemic of infectious disease (35).

## Acknowledgements

We thank Alex Pompe and Lauren McGorman from Facebook Data for Good for granting the access to Facebook Disaster Maps to conduct this study. We also thank Andrew Schroeder from Direct Relief for his input on conceptualizing the research ideas. The research carried out at the Jet Propulsion Laboratory, California Institute of Technology, was supported by the National Aeronautics and Space Administration (NASA) Land-Cover and Land-Use Change (LCLUC) Program.

# Supplemental Information

## 1. Facebook Disaster Maps (FBDM) data availability for Mendocino Complex fire

Compared to Woolsey fire, the Mendocino Complex fire posed a more complicated timeline of fire progression. Mendocino Complex consists of two separate but nearby fires, the earlier occurred, smaller River fire and the later occurred, larger Ranch fire. The relatively small perimeter of Rive fire did not activate a FBDM data generation ticket after it broke out, resulting to an incomplete time series to cover the population displacement during Mendocino Complex fire from the beginning. FBDM data were not generated until August 1, 2018, when the fire progression was out of control and rapidly claimed a large amount of area on the east side of the downwind direction. At the same time, area affected by River fire has lifted the mandatory evacuation orders, causing a divergent population change trend near River fire and the Ranch fire. Due to the limited data availability, FBDM was able to record the population displacement from the announcement of evacuation order to the end of returning process for the west side on the downwind direction of Ranch fire only. Readers can refer to Fig. 2 for FBDM data availability of different dates and regions.

An interactive version of the results can be access via below link.
https://storymaps.arcgis.com/stories/c1563816b84e4520951e1325e3fcb77b

## 2. Facebook Disaster Maps (FBDM) data specifics and accessibility

FBDM data are generated based on the location history and use of Facebook mobile application befoe and during the crisis. The data are deidentified and aggregated to 1 km grids over an area of interest every 8 hours once a data production ticket is activated in response to a significant regional or global crisis (1). The data are provided as the crisis continues until situation resettles. Retroactive data generation is possible upon request.

FBDM data include a baseline population metric to illustrate the pre-crisis Facebook usage in the area of interest. Other metrics of FBDM include crisis population, percentage of population difference, and z-score of crisis population. The z-score is calculated as follows:

$$\frac{c-\mu_{baseline}}{\max\,[\sigma_{baseline},\sigma_{min}]} \quad \text{(Eq. 1)}$$

where $c$ is the crisis population, $\mu_{baseline}$ is the mean of baseline (pre-crisis) population, $\sigma_{baseline}$ is the variance of baseline (pre-crisis) population. $\sigma_{min} \approx 0.1$ is introduced to handle the case where there is no variance in the baseline distribution (1).

To ensure the use of FBDM is constrained to uncommercial purposes and protect the digital privacy, FBDM data are distributed via Facebook GeoInsights portal for authorized users. Interested parties shall contact program managers of Facebook Data for Good (dataforgood.fb.com) and fill the application form (2) with information on the purpose of data use.

## 3. Evaluating the representativeness of Facebook Disaster Maps

We investigated the relationship between the penetration rate and the population of different age groups. Although the adjusted $R^2$ of these correlations was very low (< 0.1) (SI Fig. 1), the younger age groups showed a higher and positive correlation with the penetration rate of Facebook usage. Zip codes with a greater population born in and after 1981 showed a clearly higher correlation with Facebook usage penetration rate. Generation Alpha (born in 2017 or later) showed a similar correlation as Millennials because these two age groups are bonded together by the parental relationship. In contrast, older age groups (Baby Boomer, Silent & Greatest Generations) showed a negative correlation with the Facebook usage penetration. Such difference between age groups indicated that FBP sampled more heavily among younger age groups, though it preserves a high fitness with the total population.

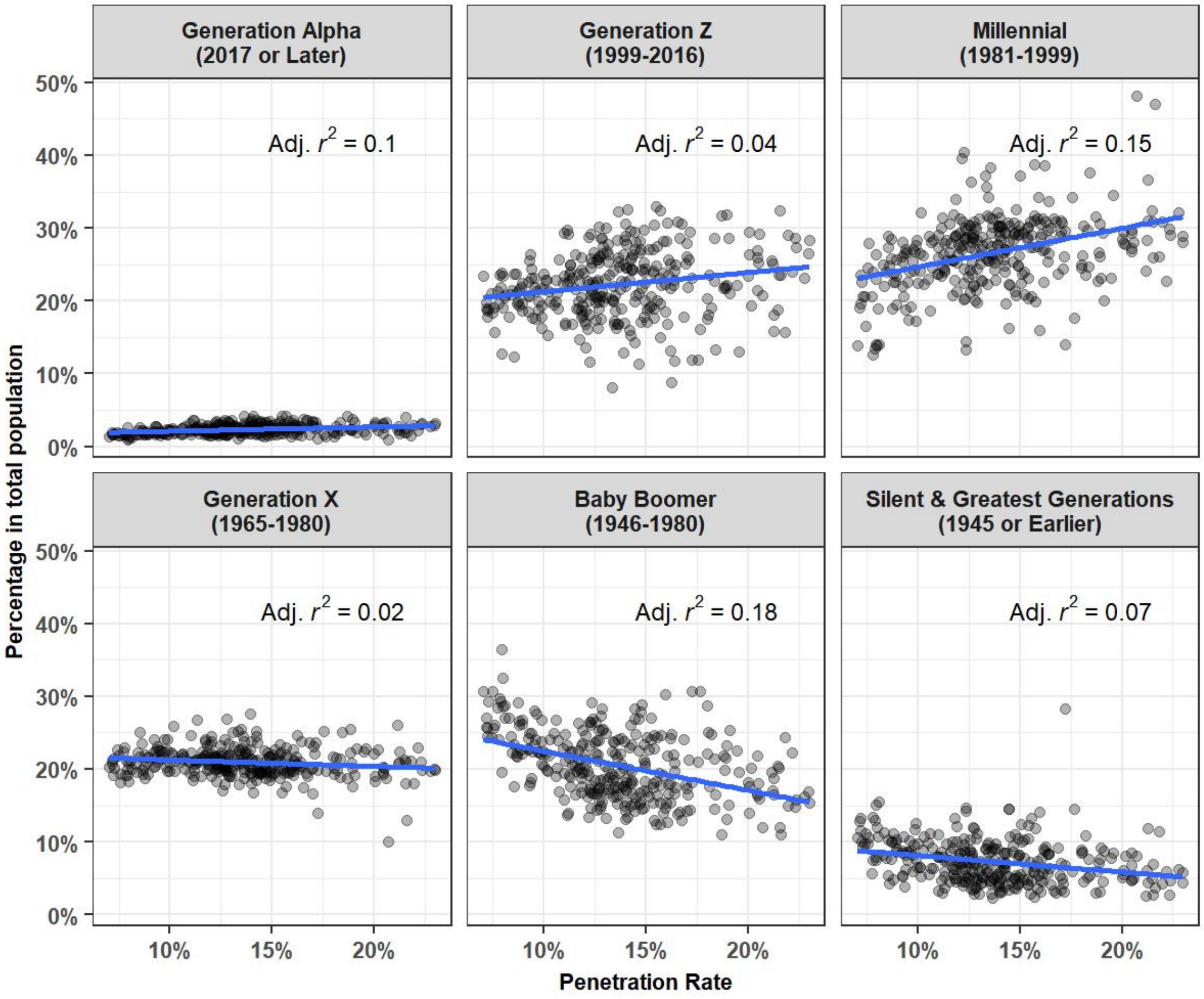

Figure S1. Penetration rate of regular Facebook users from the FBDM Woolsey fire dataset and percentage of different age groups out of the total population for each zip code. Each point represents one zip code. Blue lines indicate the linear fit between penetration rate and ACS population. Adjusted $R^2$ of the linear fit is presented for each panel.

We also found that the FBDM baseline population had a diurnal cycle consistent with the human activities. Multi-day average of penetration rate for the three sampling time of the day (1 AM, 9 AM, and 5 PM) showed that the regular Facebook usage had a diurnal cycle. Regardless of the total population dataset, penetration rate of regular Facebook users in the total population was the lowest after the midnight (1 AM). The average penetration rate peaked in the morning (9 AM) while median penetration rate peaked in the late afternoon (5 PM). The diurnal and weekly cycle indicated that FBP captured the daily human activity cycle.

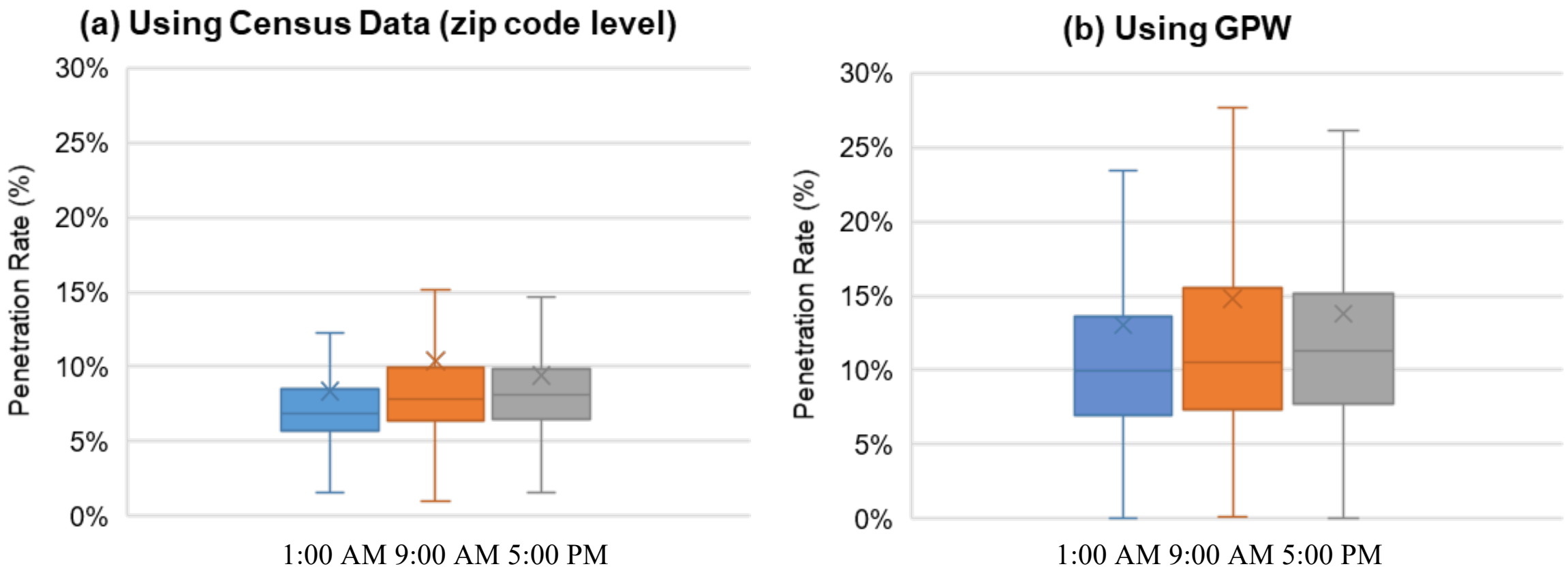

Figure S2. Multi-day average of Facebook penetration rate in Los Angeles and Ventura County, California from November 8-19, 2018 calculated using ACS 2018 population estimation at zip code level (a) and GPW (b).

**4. Population displacement shown as the population difference from pre-crisis condition**

We calculated the Facebook population difference between crisis and pre-crisis for every 1-km grid at each time step during Mendocino Complex fire and Woolsey fire and added up over the entire fire-affected area. This metric provides a big picture of the population difference from the pre-crisis population. We did not use an average of population difference as the presence of both positive and negative value makes the average around zero, which does not align with the actual population difference in the study area.

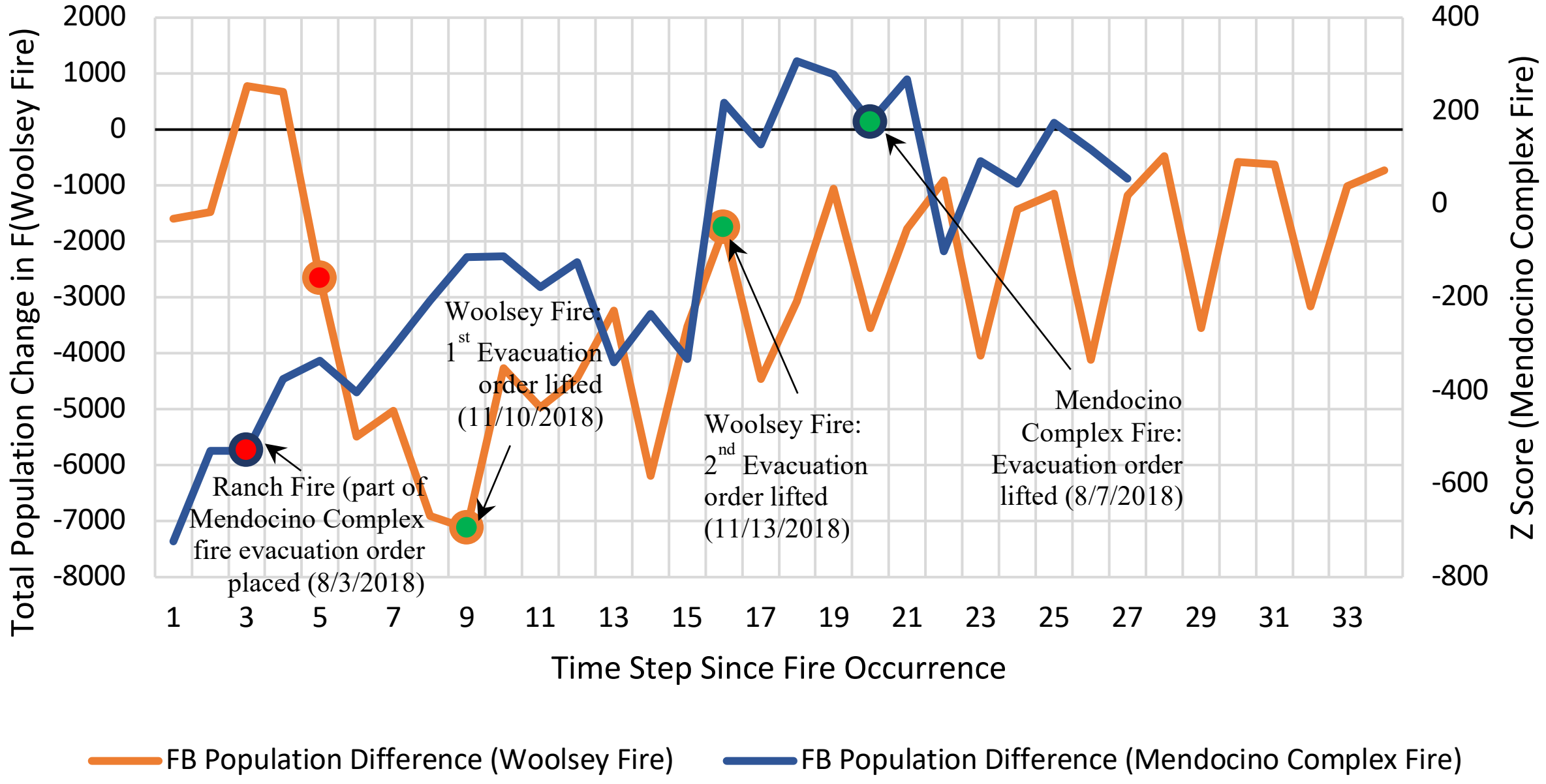

Figure S3. Total Facebook population change during Mendocino Complex fire and Woolsey fire. Red circles indicate the issue of an evacuation order. Green circles indicate evacuation order lifting. Facebook population change are presented in an 8-hour time step.

Figure S3 is provided as another format of showing the progression of population change during Mendocino Complex fire and Woolsey fire. The total change of Facebook population over the entire fire-affected area during Mendocino Complex fire and Woolsey fire is presented. Geographical boundary of the fire-affected area is consistent with what has been used to created Fig. 5 in the main text. The total Facebook population change showed a similar trend as the average of z-score for the fire-affected area (Fig. 5).

However, this metric did not capture the significant decrease of population after the evacuation order was placed for Ranch fire (the larger fire that consists of the Mendocino Complex fire), as the population was undergoing a surge in the area near River fire, the smaller fire as a part of Mendocino Complex fire, where the evacuation order was lifted not long ago. Such The complexity of evacuation order placing and lifting in the affected area of Mendocino Complex fire posed some challenge in capturing the more important signal of population change. Therefore, adopting z-score would be more useful in capturing the temporal pattern of population change during the crisis.

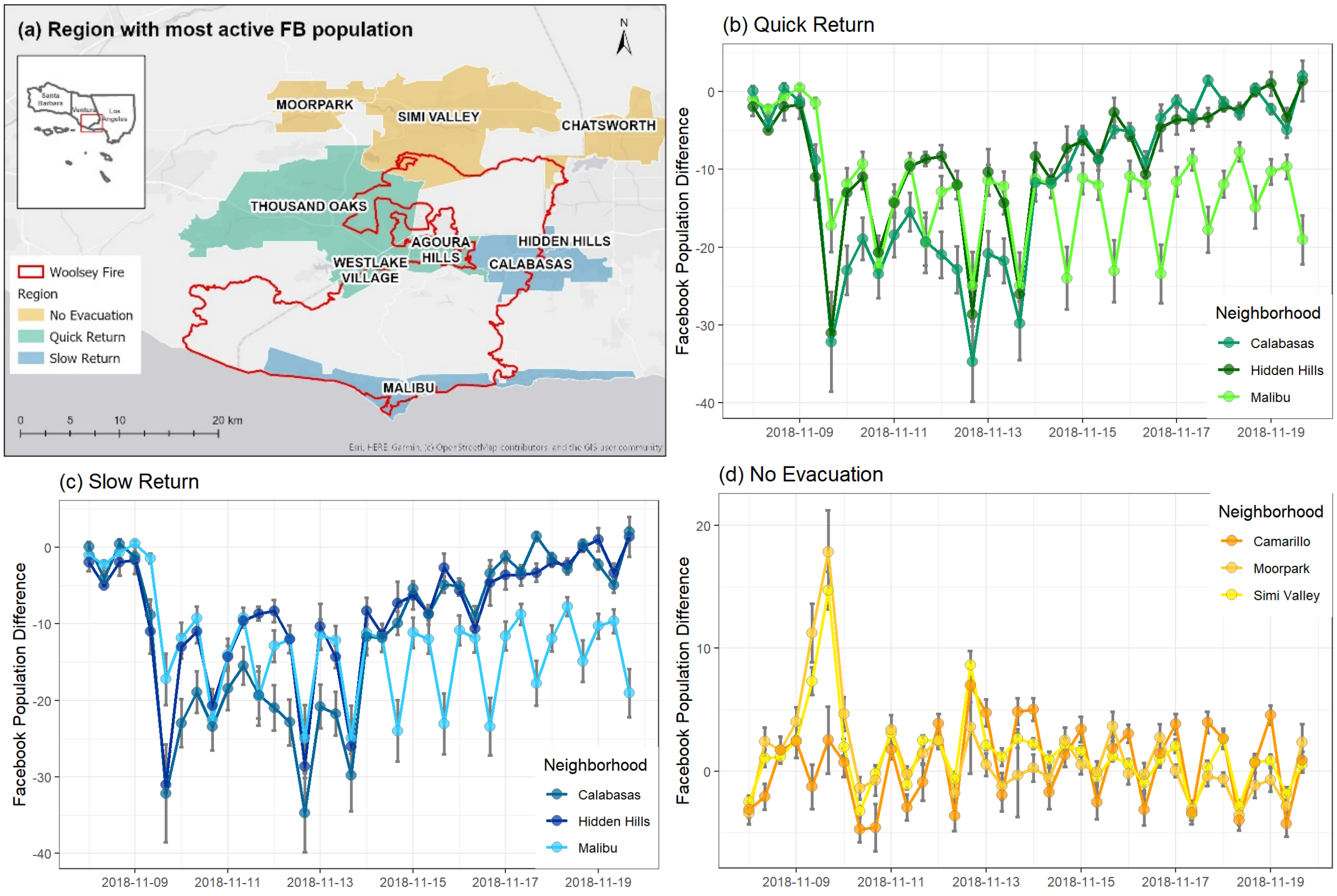

Figure S4. Facebook Disaster Maps (FBDM) population change of every 8 hours during the progression of Woolsey Fire for different locations due to the different fire progression extent. Green, blue, yellow indicated neighborhoods with quick return, slow return, and no evacuation. Line charts from panel b to d represent the temporal change of Facebook population the three types of neighborhoods mentioned above. Error bars are plotted using ± standard error.

Figure S4 is provided as another format of showing the progression of population change in three regions with different patterns of population dynamics during Woolsey fire. It can be compared with Fig. 6 in the main text, which uses z-score as the metric for plotting. The Facebook population change showed similar pattern with z-score with a more drastic change.